\documentclass[
superscriptaddress,
showpacs,
 amsmath,amssymb,
prl,
]{revtex4}

\usepackage{graphicx}
\usepackage{dcolumn}
\usepackage{natbib, textcase,bm}

\begin{document}

\title{Laminar shocks in high power laser plasma interactions.}

\author{R. A. Cairns}
\affiliation{University of St Andrews, North Haugh, St.~Andrews,
Fife, KY16 9SS, UK}
\author{R. Bingham}
\altaffiliation{Also at: SUPA, Department of Physics, University of
  Strathclyde, Glasgow, Scotland G4 ONG, UK}
\author{P. Norreys}
\altaffiliation{Also at: Physics Department, University of Oxford, Oxford OX1
  2JD, UK}
\author{R. Trines}
\affiliation{Central Laser Facility, STFC, Rutherford Appleton
Laboratory, Harwell Oxford, Didcot, OX11 0QX, UK}

\date{\today}

\begin{abstract}
We propose a theory to describe laminar ion sound structures in a
collisionless plasma. Reflection of a small fraction of the upstream
ions converts the well known ion acoustic soliton into a structure
with a steep potential gradient upstream and with downstream oscillations.
The theory provides a simple interpretation of results dating back
more than forty years but, more importantly, is shown to provide an
explanation for recent observations on laser produced plasmas
relevant to inertial fusion and to ion acceleration.
\end{abstract}


\maketitle
\renewcommand\thesection{\Roman{section}}
\section{I. Introduction }
Some recent experiments on the interaction  of high power lasers with
plasmas
have shown evidence of shock-like structures with very high electric
fields
existing over very short distances\cite{li08, ryg08}. Amendt et al
\cite{ame09} say that data
from proton radiography in inertial confinement fusion capsules
suggest the
existence of fields of more than $10^{10}$ Vm$^{-1}$ over distances
of the
order of 10-100 nm. For the non-cryogenic targets used in the experiment Amendt et al (2009) demonstrated that the experiment behaves as a quasi-collisionless classical plasma, this justifies the use of a collisionless theory especially when considering the extremely small length scales over which the electric field exist. In a more recent paper Amendt et al \cite{ame11}
suggest that barodiffusion (ie
pressure-driven diffusion) may be a possible explanation, but this
does not seem to produce very short length scales. Another relevant
recent paper is that of Haberberger et al \cite{hab12} who describe
experiments in which collisionless shocks generate high energy proton
beams
with small energy spread.

Our objective here is to show that there is a simple analytic
treatment of
collisionless shock structure in unmagnetized plasmas which can
reproduce the
essential features of these experiments and which may be useful in
predicting
the properties of shocks in collisionless plasmas. The basic method
goes back
to early studies of collisionless shocks, in particular the work of
Sagdeev
\cite{sag66}, in which it is shown that solitary wave structures can
be
described by an equation analogous to that of a particle moving in a
potential
(now usually referred to as the Sagdeev potential). The Sagdeev
potential is a
function of the  electrostatic potential $\phi$ and a solitary wave
occurs
when the Sagdeev potential has a maximum at the origin then goes
through zero
again at some finite value of $\phi.$ In terms of the particle motion
analogy
there is a homoclinic orbit in its phase space leaving the origin,
going to
the other zero of the Sagdeev potential,  then returning to the
origin over an
infinite time period. Sagdeev suggested that a shock like structure
could be
produced by introducing some damping into the system, so that the
orbit,
instead of returning to the origin ended up at the bottom of the
potential
well. A comprehensive review of analytical work describing electrostatic shocks is treated by Tidman and Krall \cite{tid71}.  Here we show that a shock structure can be produced by having a
finite
ion temperature so that some ions are reflected by the potential
maximum at
the shock. This produces the asymmetry between the upstream and
downstream
sides which destroys the familiar symmetrical ion sound solitary
wave. The
idea of reflection from the shock front has been familiar for many
years,
especially in studies of perpendicular shocks in magnetized plasmas
where the
reflected ions are turned around by the magnetic field and produce a
foot
structure (see for example the analysis of Woods \cite{woo71}). In a
collisionless unmagnetized plasma the reflected ions simply travel
upstream
unimpeded. Early observations of electrostatic shocks were made by
Taylor et al \cite{tay70} showing the kind of structure we describe,
a potential ramp followed by
downstream oscillations, at low Mach numbers. Computer simulations by
Forslund and Freidberg \cite{for71} later showed shocks, with more
complicated dissipative structures at higher Mach number.  More
recent PIC simulations by Fiuza et al \cite{fui12} also report shocks at higher Mach numbers than the ones used in
this paper.  Some work on this latter problem has been carried out by
Smirnovskii \cite{smi98, smi00} using basic ideas similar to
ours presented here.  The problem has been generalized to the relativistic case by Stockem et al \cite{sto13} Our objective is to give a more transparent
account of the theory and to relate it to the recent experimental
results mentioned above.

\bigskip
\section{II. Theory}
Consider collisionless ions flowing into a region where the potential
increases from zero to some positive value $\phi_{\max.}$ Taking the
incoming
ions to have a Maxwellian distribution with average velocity $V$ the
density where the potential is $\phi$, normalised to the initial
density of
the incoming flow is%

\begin{widetext}
\begin{equation}
	n_{i}(\phi,\phi_{\max})=\frac{1}{\sqrt{2\pi}}
	{\int\limits_{0}^{\infty}}
	\exp\left[  -\frac{\left(  \sqrt{v^{2}+2\phi}-V\right)
^{2}}{2}\right]dv
	+\frac{1}{\sqrt{2\pi}}
	{\int \limits_{0}^{\sqrt{2(\phi_{\max}-\phi)}}}
	\exp\left[  -\frac{\left(  \sqrt{v^{2}+2\phi}-V\right)
^{2}}{2}\right]dv
	\label{Ion density}
\end{equation}
\end{widetext}
with ion velocities normalised to the thermal velocity
$V_{i}=\sqrt{\kappa T_{i}}$ and the potential to $\frac{m_{i}V_{i}^{2}}{Ze}$ with $Z$ the
ion charge state, and $\kappa=k/m_{i}$ where $k$ is Boltzmann's constant
and $m_{i}$ is the ion mass. We assume that $V$ is sufficiently large
that the backward part
of the Maxwellian in the shock frame is negligible. The second term
here takes
account of particles reflected from the potential maximum. This, of
course,
cannot be chosen as an independent parameter but has to be consistent
with the
plasma dynamics, which is why it is included as an argument in
$n_{i}.$ Its
evaluation will be discussed later.

For the electrons we assume thermal equilibrium in the potential,
with the
electrons flowing to produce charge equilibrium far upstream where the
potential tends to zero, so that
\begin{equation}
	n_{e}(\phi,\phi_{\max})=Z n_{i}(0,\phi_{\max})\exp\left(\frac{\phi}{T}\right)
	\label{electron dens}%
\end{equation}
where $T=\frac{ZT_{e}}{T_{i}}$.

Poisson's equation then gives
\begin{equation}
	\frac{d^{2}\phi}{dx^{2}}=n_{e}(\phi,\phi_{\max})-Z n_{i}(\phi,\phi_{\max})
	\label{Poisson}
\end{equation}
with distances scaled to $\frac{V_{i}}{\omega_{pi}}.$ In order to find
$\phi_{\max}$ self-consistently we introduce the Sagdeev potential
$\Phi%
(\phi,\phi_{\max})$ through%
\begin{equation}
	\Phi(\phi,\phi_{\max})=\int_{0}^{\phi}\left[
Z n_{i}(\phi^{\prime},\phi_{\max})-n_{e}
(\phi^{\prime},\phi_{\max})\right]  d\phi^{\prime}\label{Sag Pot}%
\end{equation}
so that (\ref{Poisson}) becomes
\begin{equation}
	\frac{d^{2}\phi}{dx^{2}}=-\frac{\partial \Phi}{\partial\phi}%
	,\label{Poisson 2}%
\end{equation}
analogous to the equation of motion of a particle in a potential.

\begin{figure}
	 \includegraphics[width=0.5\textwidth,bb=0 0 850 650]{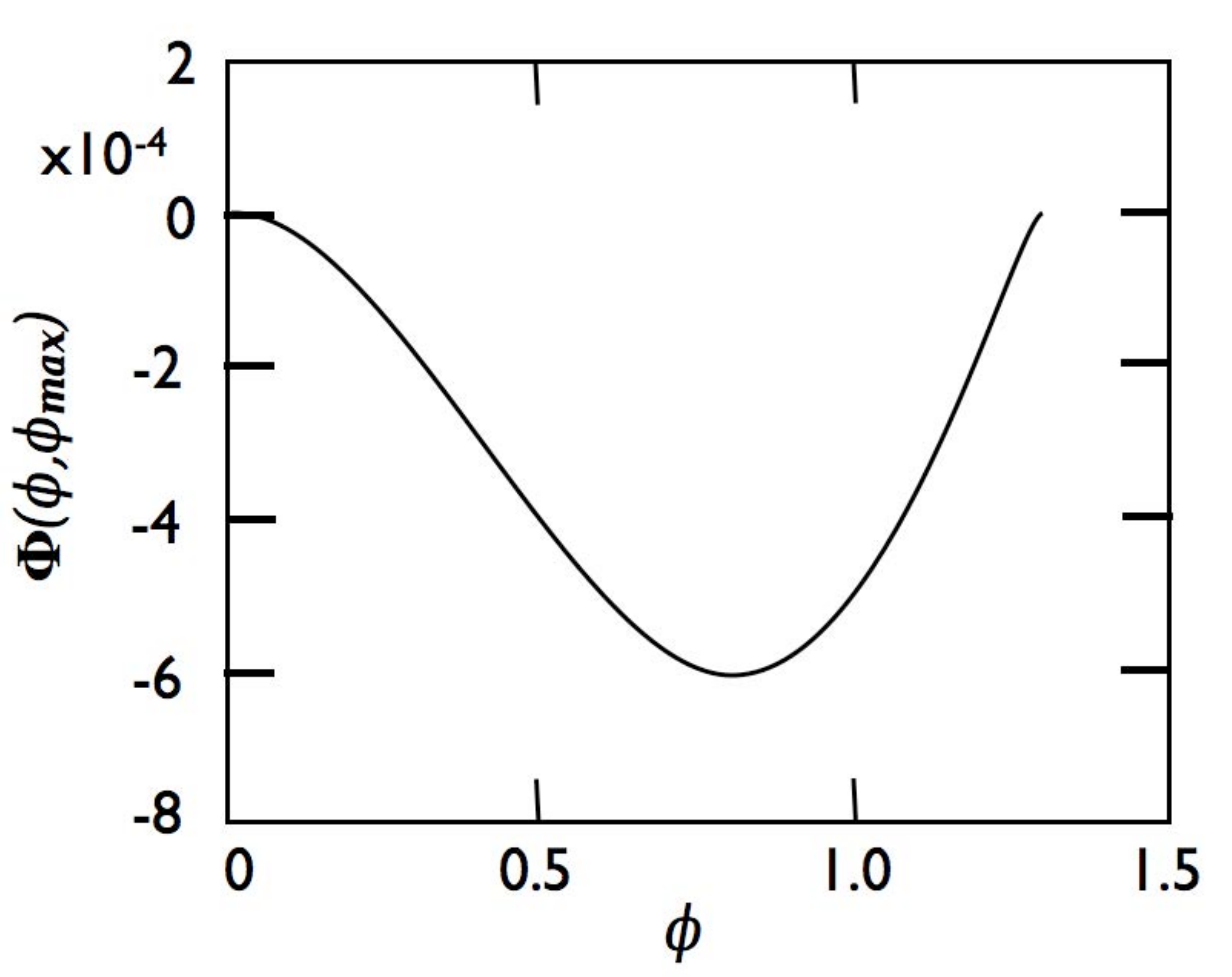}
	 \caption{The Sagdeev potential for $T$=15, $V$=4.5.}
	 \label{FIG1}
\end{figure}

The quantity $\phi_{\max}$ is still an unknown. To determine it we
note that
if the motion of a notional particle according to (\ref{Poisson 2})
starts at
$\phi=0$ then it will increase monotonically to $\phi_{\max}$ if the
Sagdeev
potential is zero at $\phi_{\max}$ and negative when $\phi$ lies
between zero and $\phi_{\max}$. So the condition that the value of
$\phi_{\max}$ be
consistent with the system dynamics is that
\begin{equation}
	\Phi(\phi_{\max},\phi_{\max})=0,\label{Max pot}%
\end{equation}
an equation which determines $\phi_{\max}.$ The dimensionless
parameters
governing the system are $V$ and $T$ and it can soon be found that
not all
combinations of these yield a system in which the Sagdeev potential
has a zero
for positive $\phi$ and is negative in the interval $(0,\phi_{\max})$
. The
complicated nature of the Sagdeev potential and its dependence on the
unknown
quantity $\phi_{\max}$ mean that we are unable to make any analytic
progress
in determining the parameter range for which a suitable potential
exists. The
observations we make on this are determined by numerical
experimentation, taking $Z=1$. It
is found that an acceptable solution only exists if the electron
temperature
is sufficiently high. We have found that a value of {\it T} of around
$15$ or
more is needed and with this value and $V=4.5$ we obtain the Sagdeev
potential
of Figure~\ref{FIG1}.%

In our normalized units the ion sound Mach number is
$\frac{V}{\sqrt{T}},$
in this case 1.162. For this value of $T$ it appears that an
acceptable
solution only exists in a narrow range of Mach numbers between about
1.13 and 1.19

If the Sagdeev potential was the same downstream of the point where
the
potential reaches its maximum then we would just get a standard
solitary wave
solution, symmetric about this maximum. However, in the downstream
region
there is no reflected component and the second term in (\ref{Ion
density}) is
absent. This changes the Sagdeev potential and in the downstream
region the
notional particle motion which determines the solution is oscillation
in a
potential well. A composite solution can be obtained by starting at
the
maximum, with zero potential gradient and integrating upstream and
downstream
with the appropriate charge densities in (\ref{Poisson}). The result
is shown in Figure 2, with
the distance $x$ in units of the ion thermal velocity divided by the
ion
plasma frequency.%

\begin{figure}
	\includegraphics[width=0.5\textwidth,bb=0 0 850 650]{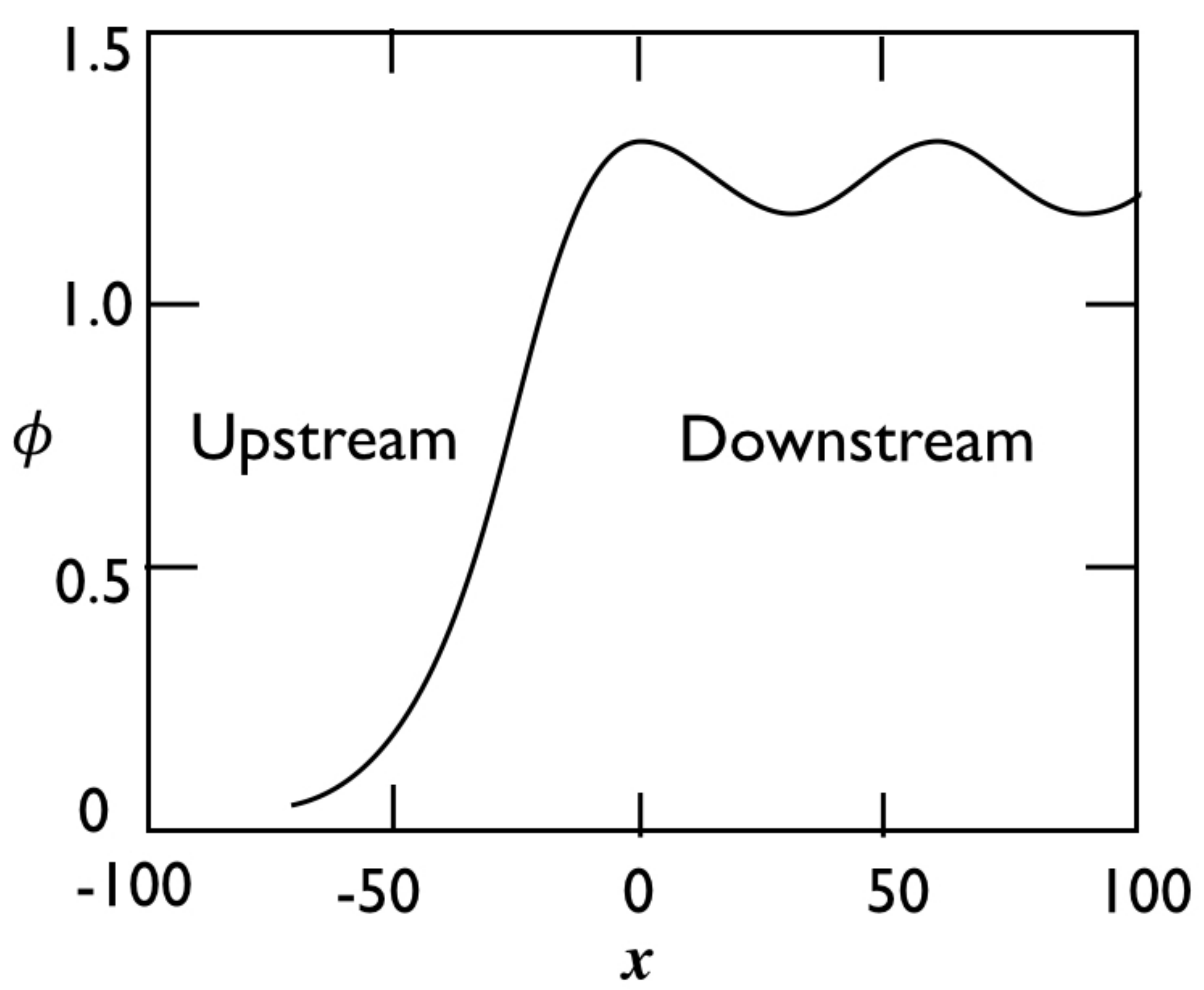}
	 \caption{The electrostatic potential for the parameters of
Figure~\ref{FIG1}}
	 \label{FIG2}
\end{figure}

If some dissipation were introduced then the notional particle would
end up at
the minimum of the potential well, corresponding to decaying
oscillations on
the downstream side and a shock like structure as described by
Sagdeev \cite{sag66}. It is worth noting that the
solution is very sensitive to small changes in the charge density. In
this
case the density as $\phi\rightarrow0$ on the upstream side goes to
1.0019, so
that very few ions are reflected, but there is nevertheless a radical
change
in the nature of the solution between the upstream and downstream
sides. For higher electron to ion temperature ratios the Mach number can be larger leading to a higher percentage of reflected ions as will be shown in the next section.

\section{III. Results}
\begin{figure}
	 \includegraphics[width=0.5\textwidth,bb=0 0 850 650]{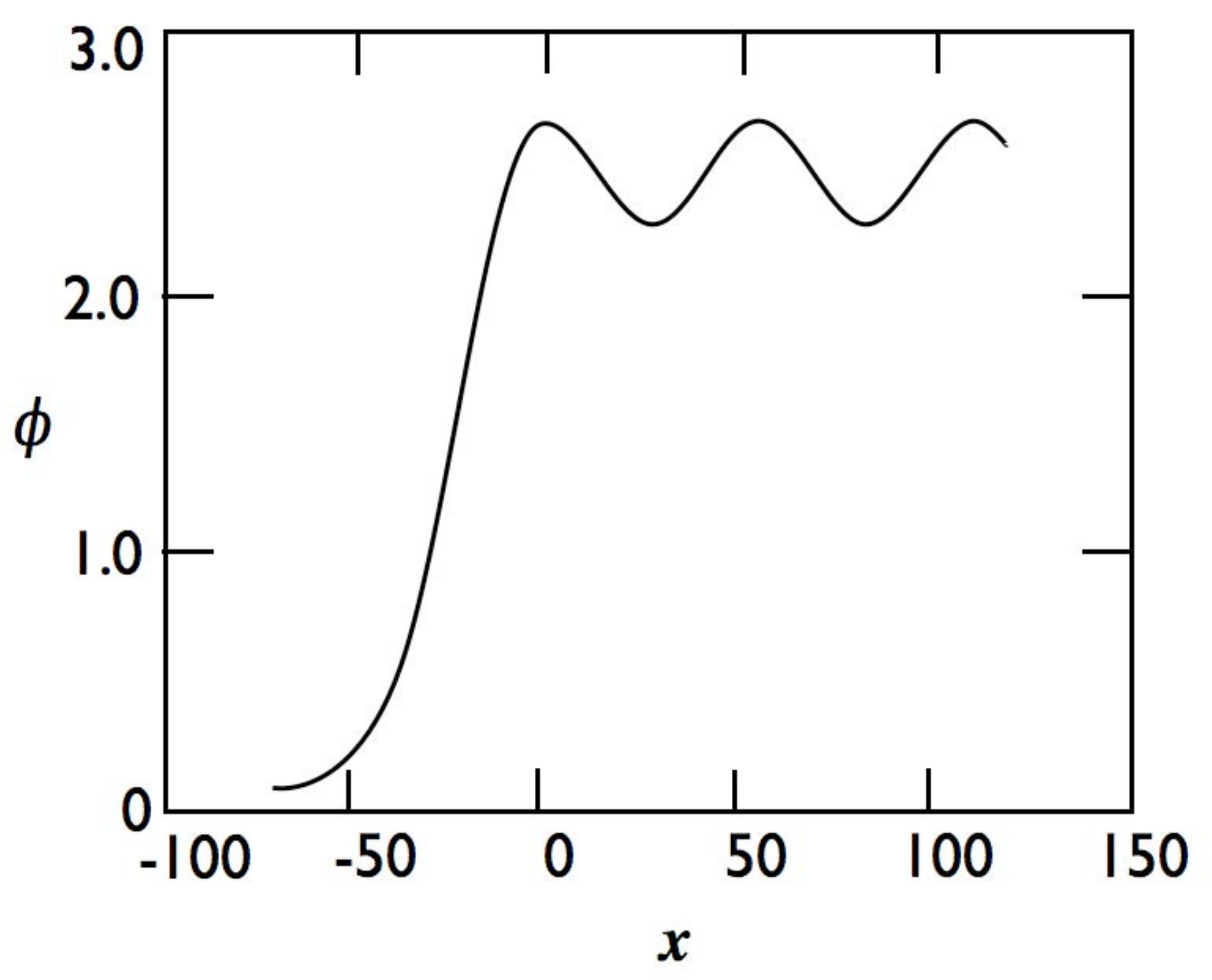}
	 \caption{The potential for the D-T plasma with $T_{e}=20$ and
$V=4.75.$}
	 \label{FIG3}
\end{figure}

To explore the possible relevance to a laser fusion pellet
compression we can do a similar
calculation with a 50/50 mixture of deuterium and tritium upstream.
With the
potential and flow speed normalised in terms of the deuterium thermal
velocity
the ion density is half the expression in (\ref{Ion density}) plus a
corresponding tritium contribution in which $\phi$ is replaced with
$\frac
{2}{3}\phi$ to take account of the higher mass. The calculation then
goes
through as before but we find that an acceptable solution only
appears to
exist for somewhat higher values of $T$. For $T=20$ and $V=4.75,$
corresponding to a Mach number of 1.06, we get the solution shown in
Figure~\ref{FIG3}. The corresponding electric field, normalized to
$\frac{m_{i}V_{i}^{2}}%
{Ze}\frac{\omega_{pi}}{V_{i}},$ is shown in Figure 4.%

\begin{figure}
	 \includegraphics[width=0.5\textwidth,bb=0 0 850 650]{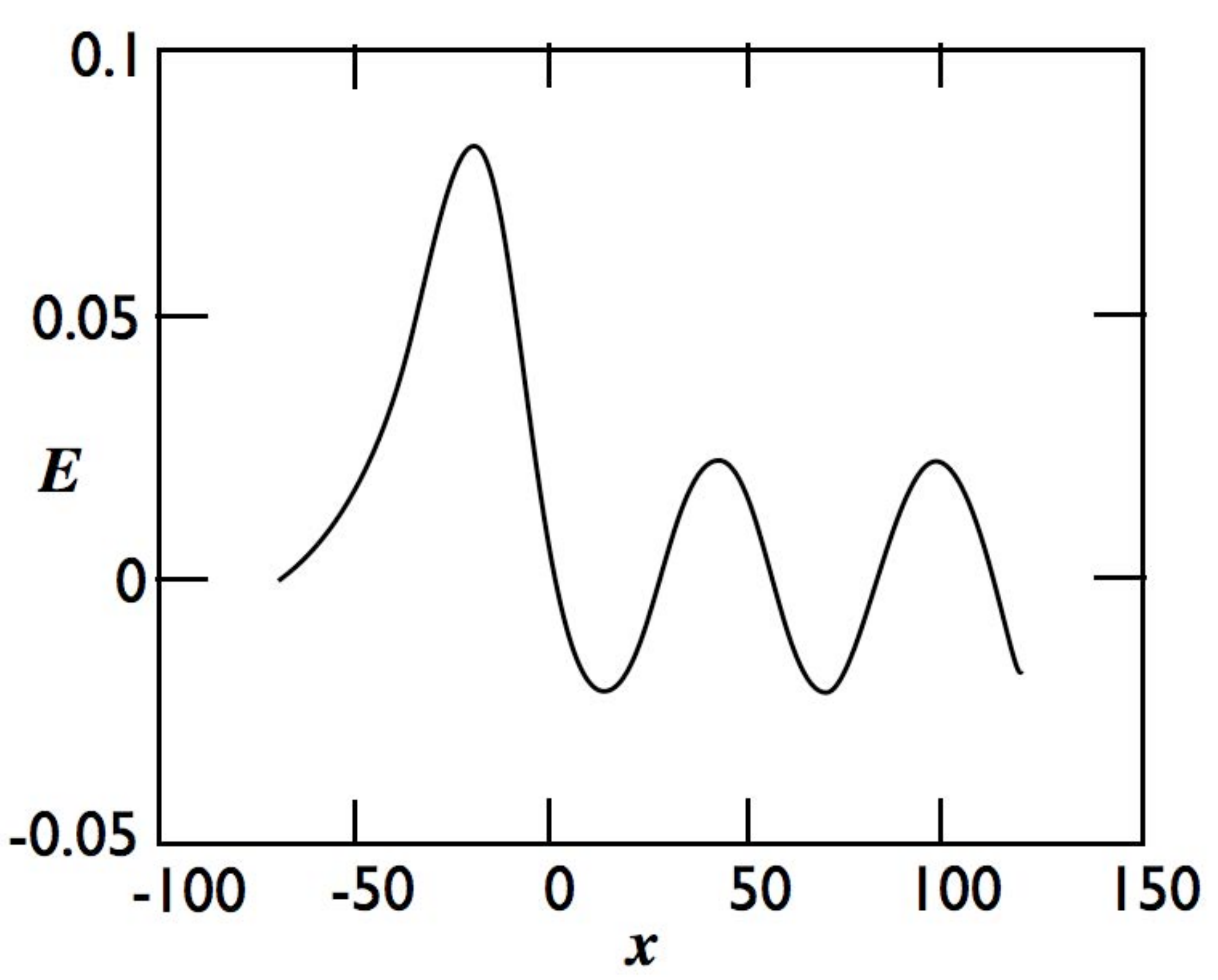}
	 \caption{The normaized electric field corresponding to the
potential of Figure~\ref{FIG3}}
	 \label{FIG4}
\end{figure}

Now let us relate these normalized values to physical parameters. If
we assume
that $Z=1,$ then  we have for the electric field and length scale
\begin{align}
		E(V/m)  & =4.27\times10^{-3}E_{norm}T_{i}(keV)^{1/2}n_{i}(m^{-3}%
)^{1/2}\label{Scaling}\\
L(m)  &
=2.34\ast10^{5}L_{norm}T_{i}(keV)^{1/2}n_{i}(m^{-3})^{-1/2}.\nonumber
\end{align}

If we look at the D-T result given above and assume an ion
temperature of 500
eV and density $10^{28}$ m$^{-3}$ then we get a peak electric field of
$2.4\times10^{10}$ V/m and, taking the normalized length of the main
potential
ramp to be 50, corresponding to a length of 83 nm. These parameters
are in striking agreement
with those quoted by Amendt et al \cite{ame09}. Note that because of
the restricted range
of Mach numbers within which these structures exist there is little
scope for
changing these values by adjusting the Mach number. Also there is a
fairly
weak dependence on density and ion temperature so that the orders of
magnitude
will remain in general agreement over a wide range of parameters. The
main
requirement is a high electron temperature compared to the ion
temperature, a
condition likely to be satisfied in high power laser plasma
interactions.  Going back to the early experiments of Taylor et
al \cite{tay70} we can take this example and, instead of scaling
to a high temperature, high density fusion
target scale it to their ion temperature of around $0.2$ eV and
density of around $10^{15}$ m$^{-3}$.  The resulting distance between
the first two potential peaks is around 5 mm, again in excellent agreement with the observations.

\begin{figure}
	 \includegraphics[width=0.5\textwidth,bb=0 0 850 650]{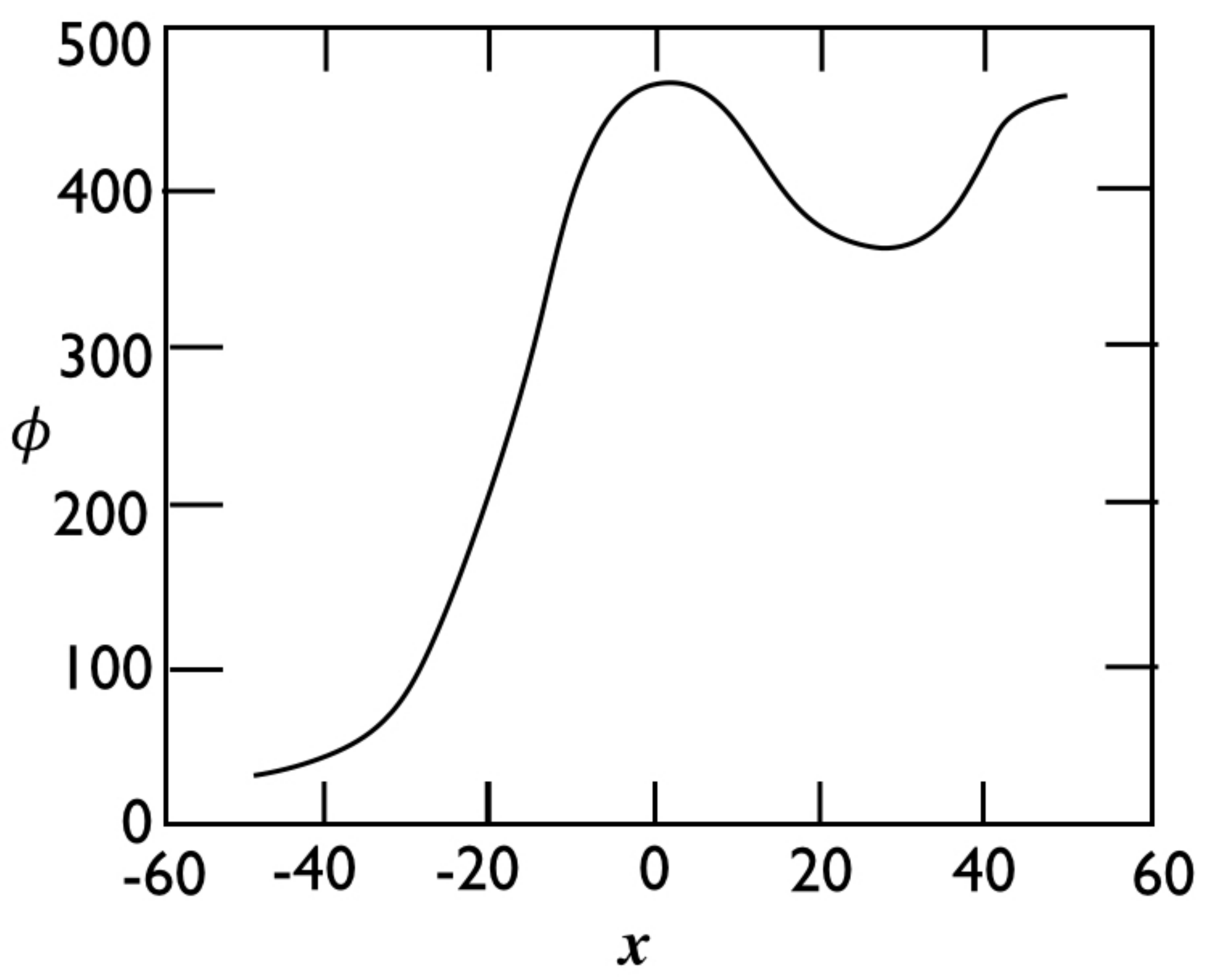}
	 \caption{The potential for $T_{e}=500$ and Mach number 1.38.}
	 \label{FIG5}
\end{figure}

Now let us look at the results of Haberberger et al \cite{hab12}
mentioned above, where
they attribute ion beams well collimated in energy to a shock wave in
an expanding plasma.
The electron temperature they find is about an MeV and we will assume
that the already
heated and expanding ions are at 2 keV, so that $T =500$. The
potential in
this case, with a Mach number of 1.38  is shown in Figure~\ref{FIG5}.%

\begin{figure}
	\includegraphics[width=0.5\textwidth,bb=0 0 850 650]{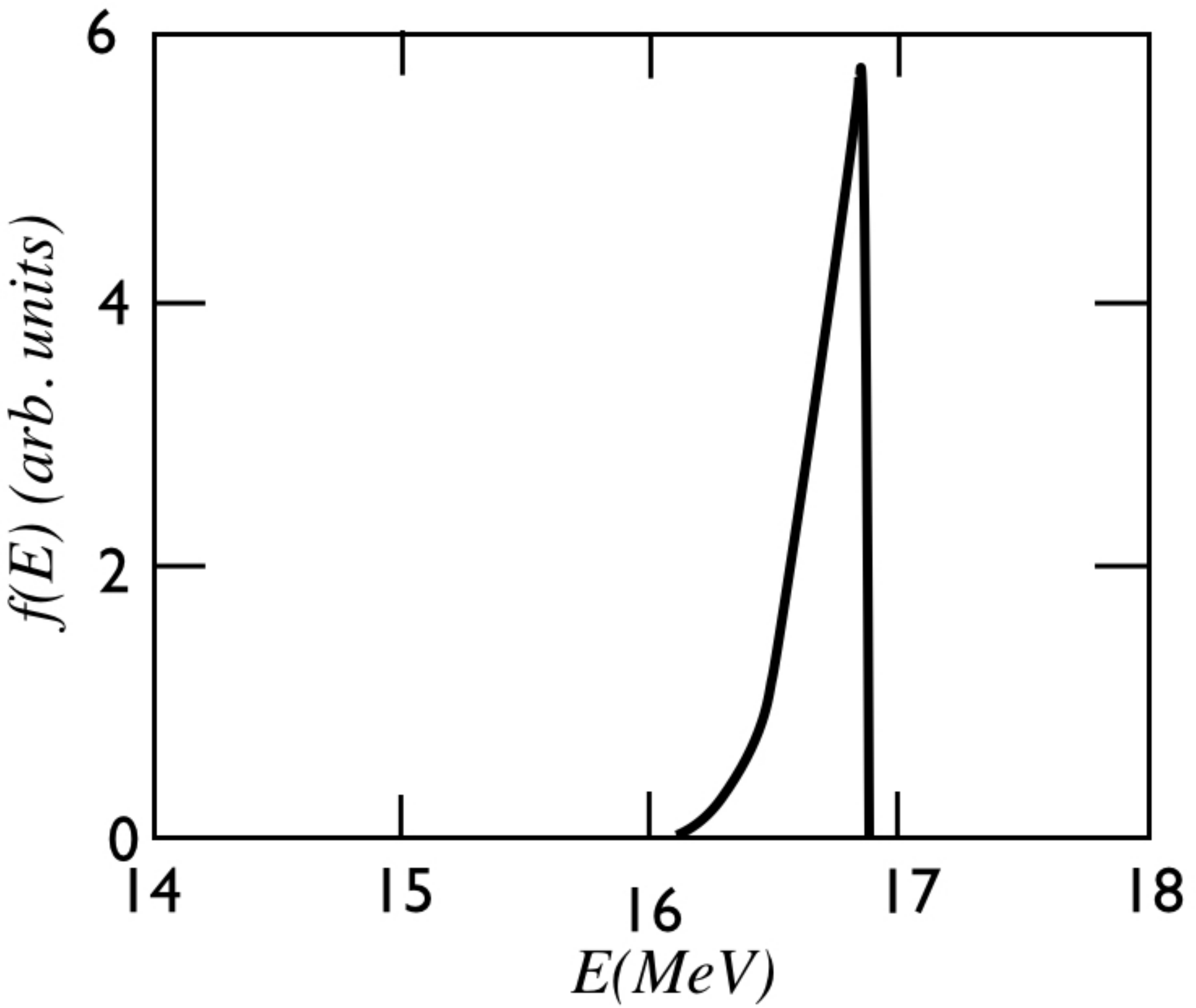}
	 \caption{The energy spectrum of reflected ions for the parameters
given in the
text.}
	 \label{FIG6}
\end{figure}

The normalized length scale is again about 50 which translates into a
physical
length of about $2$ $\mu$m if we take $n=10^{26}$ m$^{-3}$ , while
the peak
electric field is around $3.6\times10^{11}$ V/m, To compare with the
experimental results, we look at the energy spectrum of the reflected
ions.
Adding the measured  expansion velocity of $0.1c$ to the reflected ion
velocity we get the spectrum shown in Figure~\ref{FIG6}.%
This bears a striking resemblance to the experimental results, not
only in the
width of the spectrum and its energy but even in the detailed shape
with a
sharp edge on the high side. The density of reflected ions is about
36\% of
the background ion density, though this can go down
if the Mach number is reduced.  Again we should point out that
because of the
limited range of possible Mach numbers and the weak dependence of the
physical
values on the plasma parameters, we do not have a great deal of
freedom to
adjust parameters so that our results lie in the correct range. The
result
given here appears to match the experiment much better than the
computer
simulation shown in the Haberberger et al \cite{hab12} paper. One
possible explanation is
that the shock in the simulation has been launched with larger Mach
number of about 2. This is well above the limit beyond which our
laminar solutions do not
exist (around 1.4), so it may be that what is being seen is some kind
of
turbulent shock, producing a much broader spectrum of fast ions.

Going back to the early simulations of Forslund and Freidberg
\cite{for71} we see just this behaviour with the sort of structure
we describe at low Mach number but a change to a more complex
dissipative structure with many more reflected ions at higher Mach
numbers. Our results
suggest that what is seen in the experiment is the result of a low
Mach number laminar structure rather than a higher Mach number
dissipative shock.

\section{IV. Conclusion}
In conclusion, we have given a simple analytic description of laminar
shock
structures in unmagnetized plasmas and shown that the theory, despite
its
simplicity, can provide an explanation of results from important
recent
experiments on high power laser plasma interactions. It explains the
existence of very high electric fields in inertial fusion
targets and should be useful in guiding developments in the use of
lasers to produce high
quality energetic ion beams. On a more fundamental level there is a
large body
of literature on solitary waves where a heavy component, whether ions
or dust,
is assumed cold. Including thermal effects, in the way done here
could lead to
a reappraisal of these structures.

In future work we intend to investigate the conditions under which
different types of collisionless shock can be generated in laser
plasmas. A better understanding of these should help efforts to
eliminate them where they produce unwanted effects or to better
control them when they give useful effects like ion acceleration.

\section{V. Acknowledgements}
This work was supported by the UK
Engineering and Physical Science Research Council.  RB thanks STFC
Centre for Fundamental Physics for support.

\section{VI. References}

\end{document}